\documentclass[12pt ]{article}
\usepackage{jheppub_kim}
\usepackage{amsmath}
\usepackage{amssymb}
\usepackage{dcolumn}
\usepackage{bm}
\usepackage{color}
\usepackage{epsfig}
\usepackage{amsfonts}
\usepackage{graphicx}
\usepackage{subfigure}
\usepackage{multirow}
\usepackage{xcolor}
\usepackage{float}
\newcommand{\be}{\begin{equation}}
\newcommand{\ee}{\end{equation}}
\newcommand{\beq}{\begin{equation}}
\newcommand{\eeq}{\end{equation}}
\newcommand{\bea}{\begin{eqnarray}}
\newcommand{\eea}{\end{eqnarray}}
 
\def\be{\begin{equation}}
\def\ee{\end{equation}}
\def\ba{\begin{eqnarray}}
\def\ea{\end{eqnarray}}
\begin{document}

\title{Phantom BTZ Black Holes: Thermal Properties Under Perturbative Corrections}

\author[a]{Kumar Sambhav Upadhyay,}

 \author[b,c]{Sudhaker Upadhyay\footnote{Visiting Associate, Inter-University Centre for Astronomy and Astrophysics (IUCAA) Pune-411007, Maharashtra, India}} 
 \author[a]{Bhabani Prasad Mandal}
\affiliation[a]{Department of Physics,
Banaras Hindu University,  Varanasi-221005, India}
\affiliation[b]{Department of Physics, K.L.S. College, Nawada, Magadh University, Bodh Gaya,   Bihar 805110, India}
\affiliation[c]{School of Physics, Damghan University, P.O. Box 3671641167, Damghan,  Iran}

\emailAdd{kumarsambhav121@gmail.com}
\emailAdd{sudhakerupadhyay@gmail.com; sudhaker@associates.iucaa.in} 
\emailAdd{ bhabani.mandal@gmail.com; bhabani@bhu.ac.in} 

	\abstract{ 
This study investigates the thermodynamics of phantom BTZ black holes by incorporating leading-order perturbative corrections arising from small statistical fluctuations around equilibrium. Starting from the phantom BTZ black holes review, we describe a modified action in three-dimensional spacetime that includes coupling with a Maxwell or phantom field. The analysis derives the corresponding field equations and obtains exact solutions for the metric and thermodynamic quantities. The corrected entropy is computed using the steepest descent method. It is expressed in terms of the leading-order entropy and Hawking temperature. Standard thermodynamic relations yield the corrected mass, Helmholtz free energy, specific heat, and Gibbs free energy. These corrections reveal significant deviations from classical results, particularly in the small black hole regime where statistical effects become prominent. Graphical analysis shows that the corrected entropy becomes negative for sufficiently small black holes, indicating potential limitations in thermodynamic stability under perturbative corrections.
Furthermore, the influence of thermal fluctuations proves substantial for the small black holes and negligible for larger black holes. In this work, we have also calculated critical points and critical compressibility factor ($Z_{c}$) of the phantom BTZ black hole, treating it as a Van der Waals fluid. This work provides a comprehensive understanding of how statistical fluctuations modify the thermodynamics of phantom BTZ black holes. It underscores the necessity of including such corrections in realistic models of black hole thermodynamics in $(2+1)$ dimensions.
 
}
	\maketitle

\section{Introduction}

In contemporary theoretical physics, general relativity is no longer the exclusive framework for describing gravitational phenomena. One of the most compelling pathways toward a unified theory is string theory, which yields Einstein-Maxwell-dilaton gravity in the low-energy limit. Parallel to these theoretical developments, significant attention is being directed toward understanding dark energy, a fundamental component that drives the universe's accelerated expansion. A central question is whether dark energy exhibits detectable local effects at astrophysical scales. This has motivated the development of numerous effective models aimed at characterizing its nature \cite{1}. Notably, several of these models investigate the potential role of phantom fields in modeling dark energy \cite{phantom}. Many noteworthy modifications of gravity exist that are aimed at explaining dark energy. Einstein–(anti–) Maxwell-dilaton theory incorporates both Maxwell fields and a phantom dilaton field characterized by an unconventional kinetic term. In this context, a diverse range of black hole solutions has been discovered and analyzed extensively in the literature \cite{Bronnikov, Bolokhov, Gibbons, Maeda, Gurses, Rocha, Khalil, Azreg, Pacilio}.

The revelation of Hawking radiation \cite{haw}, together with Bekenstein's proposal of black hole entropy \cite{bak} and the associated thermodynamic principles, has significantly deepened our understanding of black hole dynamics. These breakthroughs have sparked strong connections between black hole theory and fundamental concepts in both thermodynamics and quantum physics. In recent decades, black hole thermodynamics has emerged as a vibrant area of study \cite{su,su1,su3,su4,su5}. For instance, the thermodynamics of Bardeen black hole  coupled with cloud of strings and  nonlinear electrodynamics is studied in Ref. \cite{rev1}. A generalized form of singular-free entropy is used to investigate the thermal properties  of higher-dimensional Reissner–Nordstr\"om black hole \cite{rev2}. The thermodynamic properties of  black holes in Einstein–Gauss–Bonnet gravity in higher dimension under the effect of exponential entropy is investigated recently \cite{rev3}. In other recent work, the thermodynamics of the 4-D Kiselev black hole is studied in the context of usual and exponential entropy \cite{rev4}.

When treated as thermal entities, black holes do not inherently comply with the second law of thermodynamics unless the notion of entropy is integrated into the analysis. Notably, it has been established that a black hole's maximum entropy corresponds proportionally to the surface area of its event horizon \cite{01, II3}. This relationship serves as the foundation for the development of the holographic principle \cite{03,04}.
However, research has shown that this maximum entropy value is not absolute; it undergoes modifications, prompting necessary adjustments to the holographic principle itself \cite{05,06}. These entropy corrections stem primarily from quantum gravitational effects and thermal fluctuations near equilibrium. Such factors become particularly influential as black holes decrease in size due to Hawking radiation.
At the leading order, these entropy corrections are known to exhibit a logarithmic form \cite{II5}. Recent studies have extended this understanding across various black hole models, including quasitopological black holes \cite{15}, charged and rotating black holes \cite{16}, those governed by $f(R)$ gravity \cite{18}, charged massive black holes \cite{17},   black branes \cite{mir}, and Horava-Lifshitz black holes \cite{19}.
In literature, there exist many important effects  of perturbative and non-perturbative  correction of entropy on the thermal properties of various kind of black holes. But the effects of such  correction on the thermodynamics as well as the stability and phase transition of phantom BTZ black holes are not studied yet. We would like to take this opportunity  to bridge this gap. This is the motivation of present investigation.

This paper reviews the work on phantom BTZ black holes \cite{Eslam_Panah_2024}. We present the action for a three-dimensional gravity theory coupled to an electromagnetic or phantom field. Then we find the expression for the corresponding field equations. Then, the static circularly symmetric solution representing a phantom BTZ black hole is obtained by solving these equations for a specific metric ansatz. The characteristics of the solution, such as the event horizon and the expressions for mass, Hawking temperature, electric charge, and potential, are studied in detail, with the mass computed using the Ashtekar-Magnon-Das approach \cite{ashtekar1984asymptotically,ashtekar2000asymptotically}. Calculating expressions of these parameters is crucial in our further work on the corrected thermodynamics of phantom BTZ black holes. In the next section, we consider the quantum-corrected thermodynamics by incorporating first-order corrections to the entropy arising from statistical fluctuations around equilibrium in the canonical ensemble. The corrected entropy is expressed in terms of the uncorrected entropy and Hawking temperature, and its behavior is analyzed. Subsequently, we derive the corrected mass using the first law of thermodynamics and analyze its dependence on the black hole horizon radius under different cosmological constant values for both Maxwell and phantom cases. Following this, we evaluate the first-order corrected thermodynamic potentials, namely Helmholtz free energy, specific heat, and Gibbs free energy, and discuss their behavior with and without thermal fluctuations. Throughout, we emphasize the contrasting thermodynamic behavior of phantom and Maxwell BTZ black holes, especially in the regime of small black hole sizes where the perturbative corrections play a significant role.
 
 The paper is presented in the following way. In Sec. \ref{sec2}, we review preliminaries about the phantom BTZ black hole. In Sec. \ref{sec3}, we 
 consider the small statistical thermal fluctuation around the equilibrium
 thermodynamics, which attributes the correction in the entropy of this black hole. In Sec. \ref{sec4}, we calculate the corrected mass of the black hole following the first law of thermodynamics due to the correction in entropy.
 In  Sec. \ref{sec5}, we study the correction in the other important thermal entities 
 due to small statistical thermal fluctuations. We emphasize the effects of thermal correction on the stability of the black hole by calculating specific heat in section \ref{sec6}. We study the $P-V$ criticality and compressibility factor for the phantom BTZ black hole by considering the black hole as a fluid 
in section \ref{sec7}. Finally, we conclude this work in section \ref{sec8}. 

\section{Phantom BTZ Black holes in $(2+1)$ spacetime dimensions}\label{sec2}

This section reviews the work on phantom BTZ black holes in $(2+1)$ spacetime dimensions \cite{Eslam_Panah_2024}.
The action for this theory in three-dimensional spacetime can be written as 
\begin{equation}
 I = \frac{1}{2 \kappa^2} \int_{\partial M} d^3x \, \sqrt{-g} \, \left[R - 2\Lambda + \eta F \right],
 \label{BTZ_ac}
\end{equation}

where $R$ is the Ricci scalar curvature and $\Lambda$ is the cosmological constant.{The third term characterizes the interaction with the Maxwell field when $\eta = 1$, whereas $\eta = -1$ corresponds to a coupling with a spin-1 phantom field. } 
 {The Maxwell invariant is defined as $F = F_{\mu\nu} F^{\mu\nu}$, where $F_{\mu\nu} = \partial_{\mu} A_{\nu} - \partial_{\nu} A_{\mu}$ represents the electromagnetic field tensor, and $A_{\mu}$ is the gauge potential. Additionally, $\kappa^2 = 8\pi G$, where $G$ denotes the Newtonian gravitational constant. We adopt the natural units $G = c = 1$ in this work. The quantity $g = \det(g_{\mu\nu})$ refers to the determinant of the metric tensor $g_{\mu\nu}$.}

Now, varying the  action given by Eq.  (\ref{BTZ_ac}) with respect to the gauge field $A_{\mu}$ and the gravitational field $g_{\mu\nu}$, the following field equations are obtained 
\begin{eqnarray}
G_{\mu\nu} + \Lambda g_{\mu\nu} &=& 2\eta\left( \frac{1}{4} g_{\mu\nu} F - F_{\mu}^{\ \alpha}F_{\nu\alpha}\right), \label{eq:G_munu} \\
\partial_{\mu}\left( \sqrt{-g} F^{\mu\nu}\right) &=& 0, \label{eq:F_Maxwell}
\end{eqnarray}

where $G_{\mu\nu}$ is the Einstein tensor.

Now, consider a three-dimensional static spacetime described by the metric form  given below 
\begin{equation}
 ds^{2} = -g(r) dt^2 + \frac{dr^2}{g(r)} + r^2 d\phi^2,
 \label{metric_BTZ}
\end{equation}
where $g(r)$ is the metric function, we must find this expression. 

The radial electric field $h(r)$ is related to the gauge potential as 
\begin{equation}
 A _\mu = h(r) \, \delta_ \mu^t = (h(r),0,0) .
 \label{A_potential}
\end{equation}
where $ h(r)$ has the following form:
\begin{equation}
 h(r) = -q \ln\left(\frac{r}{l}\right) ,
 \label{h(r)}
\end{equation}
here $q$ is the integration constant associated with the electric charge. An arbitrary constant, $l$, with the length dimension, is introduced to make logarithmic arguments dimensionless. { Using equation (\ref{h(r)}), the electromagnetic field tensor can be expressed as follows,}
\begin{equation}
  F_{tr} = \partial_t A_t - \partial_r A_t = \frac{q}{r}.
  \label{F_tr}
  \end{equation}
Putting Eq. (\ref{F_tr}) into Eq. (\ref{eq:G_munu}), we get 
\begin{eqnarray}
G_{{tt}} &=& G_{{rr}} = r g'(r) + 2 \Lambda r^2 - 2 \eta q^2 ,
\label{eq_tt}\\
G_{\mathrm{\phi \phi}} &=&  r^2 g''(r) + 2 \Lambda r^2 + 2 \eta q^2. 
 \label{eq_phi}
\end{eqnarray}
With the aid of equations (\ref{eq_tt}) and (\ref{eq_phi}), the metric function is obtained in the following manner 
\begin{equation}
  g(r) = -m_{0} - \Lambda r^2 + 2 \eta q^2 \ln\left(\frac{r}{l}\right),
    \label{metric_function}
\end{equation}
where $m_0$ and $q$ are constants related to the mass of the black hole and electric charge, respectively.

Also, phantom BTZ black holes exist for three cases of the cosmological constant ($\Lambda > 0$, $\Lambda < 0$, and $\Lambda = 0$). Each case has an event horizon for BTZ black holes in the presence of a phantom field.

{By imposing the condition $g(r) = 0$, the mass parameter $m_0$, associated with the total mass of the black hole, can be expressed as a function of the event horizon radius $r_{+}$, the cosmological constant $\Lambda$, and the charge $q$, as follows:} 
\begin{equation}
 m_{0} = 2 \eta q^2 \ln\left(\frac{r_{+}}{l}\right) - \Lambda r^2_{+}.
 \label{mass_parameter}
\end{equation}
The Ashtekar-Magnon-Das (AMD) approach \cite{ashtekar1984asymptotically,ashtekar2000asymptotically} is used to compute  the total mass of the black hole, and the following expression is obtained as
\begin{equation}
  M = \frac{m_{0}}{8}  = \frac{\eta q^2}{4} \ln\left(\frac{r_{+}}{l}\right) - \frac{\Lambda r^2_{+}}{8}.   
\end{equation}
 The surface gravity for the mentioned spacetime metric (\ref{metric_BTZ}) can be computed as:
\begin{equation}
  \kappa = \left.\frac{g'(r)}{2}\right|_{r = r_{+}}=\frac{\eta q^2}{r_{+}} - \Lambda r_{+}.
  \label{surface_gravity}
\end{equation}
Using standard definition, this leads to the Hawking temperature  $T_{H}$ of phantom BTZ black holes     as
\begin{equation}
 T_{H} = \frac{\kappa}{2 \pi}=\frac{\eta q^2}{2 \pi r_{+}} - \frac{\Lambda r_{+}}{2 \pi}.
 \label{BTZ_Temp}
\end{equation}
{The electric charge of the phantom BTZ black hole can be determined by applying Gauss's law, yielding the following expression:}
\begin{equation}
Q = \left. \int_{0}^{2\pi} F_{tr}(r) \sqrt{g} \, d\phi \right|_{r = r_+} = \frac{q}{2}.\label{Charge_Q}
\end{equation}
 Using electromagnetic field tensor $F_{\mu\nu}$, we can find the nonzero component of the gauge potential, which is $ A_t = - \int F_{{tr}}(r) dr$.
{Accordingly, the electric potential ($U$) at the event horizon, relative to the reference point $\left( r \to \infty \right)$, is given by}
\begin{equation}
 U = -  \int_{r_{+}}^{+\infty} F_{{tr}}(r)  \, dr  = q\ln\left(\frac{r_{+}}{l}\right). 
 \label{Potential_U}
\end{equation}

\section{First order corrected entropy of phantom BTZ black holes}\label{sec3}
This section considers the canonical ensemble of phantom BTZ black holes. Then, we study the effects of perturbative corrections to the thermodynamic entropy of the phantom BTZ black holes when small statistical fluctuations around equilibrium are considered. To start the analysis, let us first define the density of states with fixed energy as
        \begin{equation}
         \rho(E) = \frac{1}{2\pi i}\int_{c - i\infty}^{c + i\infty} e^{S(\beta)} \, d\beta,
         \label{Rho_E}
        \end{equation}
where $S(\beta)$ refers to  the exact entropy as a function of temperature $ T = \frac{1}{\beta}$.  The aforementioned complex integral can be solved by the method of steepest descent around the saddle point $\beta_0$ such that $\left(\frac{\partial S(\beta)}{\partial \beta} \right)_{\beta = \beta_0 } = 0$. We assume the phantom BTZ black hole is in equilibrium at Hawking temperature $T_{H}$. Now, by expanding the exact entropy using Taylor expansion around the saddle point $\beta = \beta_0$, we have
        \begin{equation}
          S(\beta) = S_0 + \frac{1}{2} (\beta - \beta_0)^2 \left( \frac{\partial^2 S(\beta)}{\partial \beta^2} \right)_{\beta = \beta_0} + \text{( higher order terms)},
          \label{S_beta}  
        \end{equation}
        where $S_0 = S(\beta_0)$ represented the uncorrected zeroth- order entropy. Now by inserting this value of $S(\beta)$ (\ref{S_beta}) into (\ref{Rho_E}) and performing integral by choosing $c = \beta_0$ for positive $\left( \frac{\partial^2 S(\beta)}{\partial \beta^2} \right)_{\beta = \beta_0}$ leads to \cite{II5}
        \begin{equation}
         \rho(E) = \frac{e^{S_0}}{\sqrt{2\pi \left( \frac{\partial^2 S(\beta)}{\partial \beta^2} \right)_{\beta = \beta_0}}}.  
        \end{equation}
        The logarithm of the above density of states gives the corrected canonical entropy at equilibrium as
        \begin{equation}
         S_{c} = S_0 - \frac{1}{2} \ln\left( \frac{\partial^2 S(\beta)}{\partial \beta^2} \right)_{\beta = \beta_0} + \text{(sub leading terms)}.  
        \end{equation}
       Now, to identify the effect of this correction term on other thermodynamical quantities, we label the $1/2$ factor in the R.H.S. of above equation by $\alpha$ and call this parameter as correction parameter.    
 Therefore, the generic expression for leading order corrections to the Bekenstein-Hawking entropy formula resulting from minute statistical fluctuations at the equilibrium of a black hole can be calculated as \cite{II5}
\begin{equation}
 S_{c} = S_0 -\alpha \ln(S_0 T_H^2),
 \label{S_C}
\end{equation}
where $\alpha$, a correction parameter arising from small statistical thermal fluctuations around the equilibrium of the black hole, takes a value equal to $\frac{1}{2}$ when thermal fluctuations exist and vanishes when we do not consider thermal fluctuations for our system at equilibrium, and $T_H$ is the Hawking temperature.

Now, to obtain the uncorrected entropy of the phantom BTZ black hole, we can use the area law, which is given by
\begin{equation}
  S_0 = \frac{A}{4},  
\end{equation}
where $A$ is the event horizon area and is defined by
\begin{equation}
 A = \int_{0}^{2 \pi} \sqrt{g_{\phi\phi}} \,  d\phi\Bigg|_{r = r_{+}} = 2\pi r_{+},  
\end{equation}
as $g_{\phi\phi} = r^2$. Therefore, the uncorrected entropy of the BTZ black holes in the presence of the phantom field can be expressed as \cite{Eslam_Panah_2024}:
\begin{equation}
  S_0 = \frac{\pi r_{+}}{2}.
  \label{BTZ_entr}
\end{equation}
 Since the Hawking temperature of the phantom BTZ Black hole is given by Eq.  (\ref{BTZ_Temp}), we also consider that our black hole system remains in thermal equilibrium with its surroundings without thermal fluctuations.
Therefore, the corrected entropy $S_{c}$ due to small statistical thermal fluctuations around the equilibrium of the BTZ black hole is given by using Eq. (\ref{S_C}) as 
\begin{equation}
S_{c} = \frac{\pi r_{+}}{2} - \alpha \ln\left[\frac{\pi r_{+}}{2} \left(\frac{ \eta \, q^2}{2 \pi \, r_{+}} - \frac{\Lambda \, r_+}{2 \pi}\right)^2\right],\label{corr}
\end{equation}
where $ S_0$  is given by Eq.  (\ref{BTZ_entr}), Hawking temperature $T_{H}$ is given by Eq.  (\ref{BTZ_Temp}), and $\alpha$ (correction parameter) represents the effect of the thermal fluctuations on the entropy of our black hole system.
\begin{figure}[htbp]
\centering
\begin{tabular}{cc} 
    \includegraphics[width=0.45\textwidth]{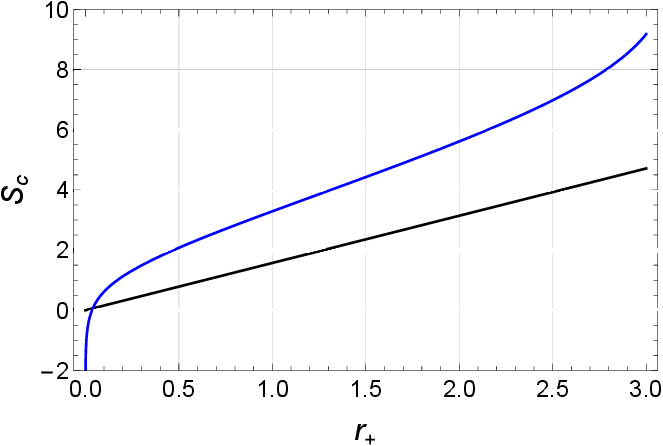}
    \includegraphics[width=0.45\textwidth]{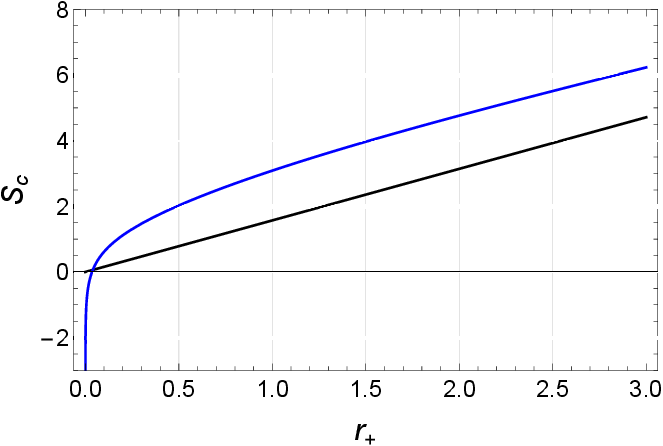}
    \end{tabular} 
    \caption{Entropy vs phantom BTZ Black hole event horizon radius. Here, $\alpha=0$ is represented by the black curve, and the blue curve represents $\alpha=0.5$. Left panel: $\eta=1$ (phantom case). Right panel: $\eta=-1$ (Maxwell case).}
    \label{fig1}
\end{figure}

From Fig. \ref{fig1}, we see that the leading-order corrected entropy after incorporating minute statistical thermal fluctuations around the equilibrium of the black hole is an asymptotically negative value for the tiny black holes in both the phantom field and the Maxwell field cases. {The appearance of negative entropy is unphysical and this implies that for very tiny black holes the perturbative logarithmic corrections to entropy have broken down \cite{su1}. Therefore, we inferred that the negative entropy values in the context of phantom BTZ black holes are physically meaningless and hence forbidden}. However, corrected entropy attains more positive values as the size of the black hole increases {and is an increasing function of event horizon radius.}

\section{Corrected mass of phantom BTZ black holes}\label{sec4}
From the first law of thermodynamics:
\begin{equation}
 dM_{c} = T_{H}dS_{c} + \eta U dQ ,
 \label{BTZ_M}
\end{equation}
where $S_{c}$ is given by eqn. (\ref{corr}). The electric charge $Q$ of a black hole is given by the equation (\ref{Charge_Q}).

\begin{figure}[htbp]
\centering
\begin{tabular}{cc} 
    \includegraphics[width=0.45\textwidth]{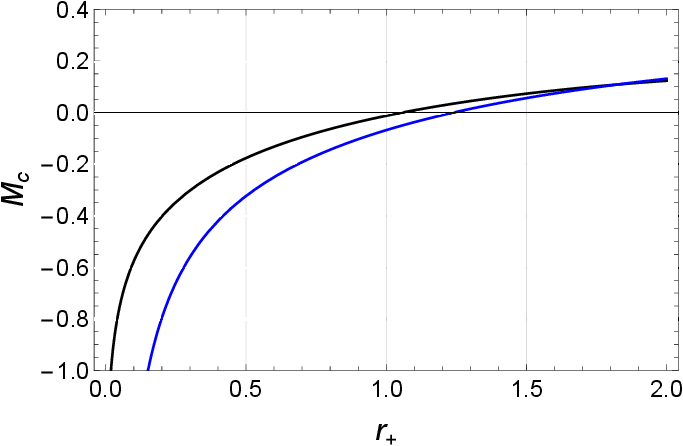}
    \includegraphics[width=0.45\textwidth]{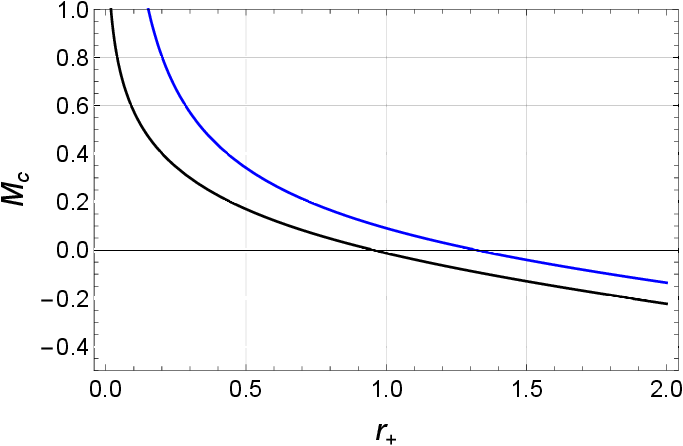}
  \end{tabular}
  \caption{Mass vs $r_{+}$ of BTZ black hole with $\Lambda = 0.1$. The black curve represents $\alpha = 0$, and the blue curve represents $\alpha = 0.5$. Left panel:  phantom case ($\eta = 1$). Right panel:   Maxwell case ($\eta =- 1$). 
      }
    \label{fig2}
\end{figure}

\begin{figure}[htbp]
\centering
\begin{tabular}{cc} 
    \includegraphics[width=0.45\textwidth]{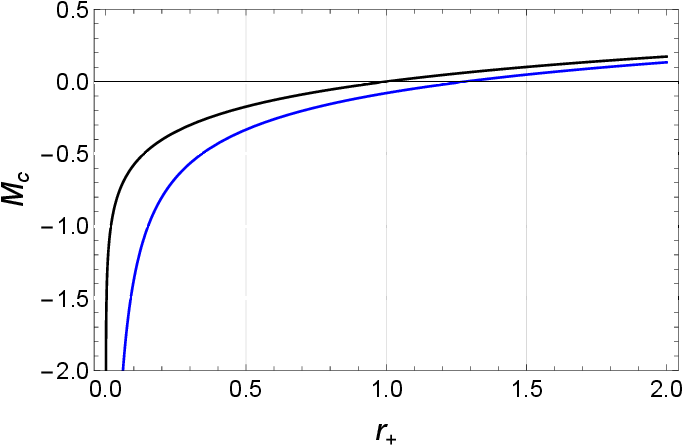}
    \includegraphics[width=0.45\textwidth]{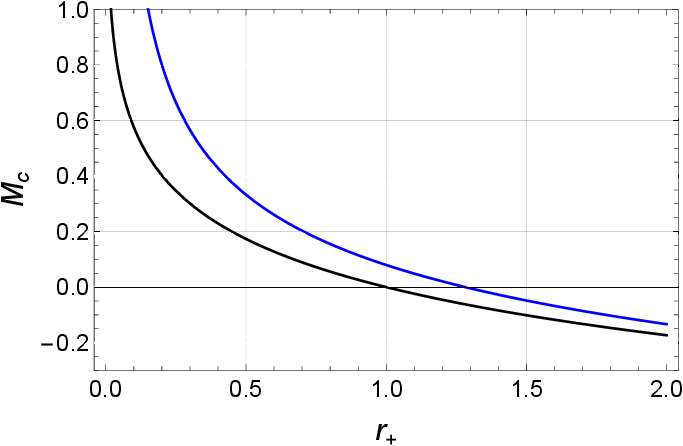}
  \end{tabular}
  \caption{Mass vs $r_{+}$ of BTZ black hole with $\Lambda = 0$. The black curve represents $\alpha = 0$, and the blue curve represents $\alpha = 0.5$. Left panel:  phantom case ($\eta = 1$). Right panel:   Maxwell case ($\eta =- 1$). 
      }
    \label{fig3}
\end{figure}

\begin{figure}[htbp]
\centering
\begin{tabular}{cc} 
    \includegraphics[width=0.45\textwidth]{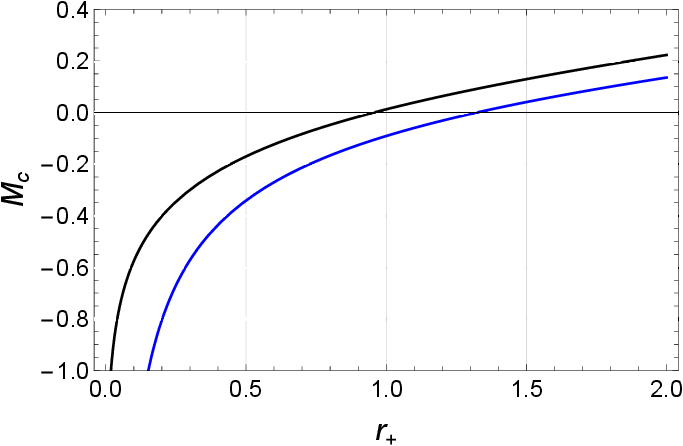}
    \includegraphics[width=0.45\textwidth]{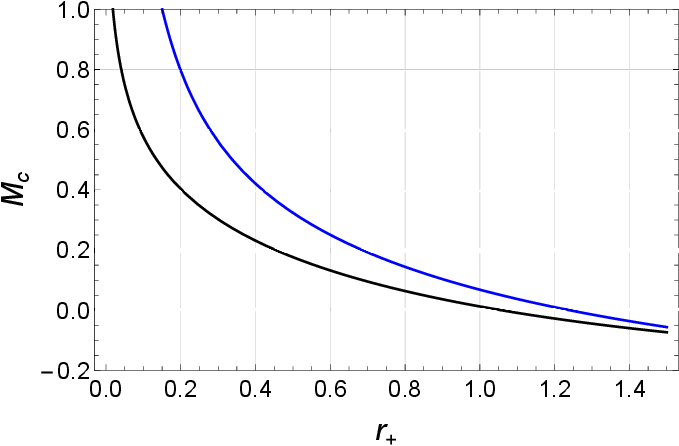}
  \end{tabular}
  \caption{Mass vs $r_{+}$ of BTZ black hole with $\Lambda = -0.1$. The black curve represents $\alpha = 0$, and the blue curve represents $\alpha = 0.5$. Left panel:  phantom case ($\eta = 1$). Right panel:   Maxwell case ($\eta =- 1$)}
    \label{fig4}
\end{figure}

{ From Eq. (\ref{BTZ_M}) we also see that for minimal horizon radius, i.e, $r_{+} \to 0$, the first law of thermodynamics breaks down completely as the product $T_{H}dS_{c}$ diverges to infinity. At $r_{+} \to 0$ we enter into the region of Planck length where the semiclassical thermodynamics, including the first law, is inapplicable; therefore, here, a quantum gravity approach becomes necessary, as the semiclassical description breaks down near the Planck scale.
Therefore, at $r_{+} \to 0$ we exit the realm of classical thermodynamics entirely and enter into quantum gravity.  } 
Since $Q=\frac{q}{2}$, where $q$ is related to the electric charge, is given by Eq. (\ref{h(r)}), which is a constant; therefore, there is no additional term corresponding to potential as $dQ = 0$.
 Upon integration the expression (\ref{BTZ_M}) leads to  
\begin{equation}
 M_c = \int T_{H} \, dS_c,   
\end{equation}
After exploiting the values of $T$ from (\ref{BTZ_Temp}) and $S_{c}$ from (\ref{corr}), the above expression simplifies to the following explicit expression:
\begin{equation}
  M_c = \frac{\eta q^2}{4} \ln\left(\frac{r_{+}}{l}\right) - \frac{\Lambda r^2_{+}}{8} + \alpha\left(-\frac{\eta q^2}{2 \pi r_{+}} + \frac{3 \Lambda r_{+}}{2 \pi}\right).
  \label{M_corrected}
\end{equation}
Now, we plot this expression for different cases of cosmological constants in both the phantom and Maxwell cases. We consider cosmological constant positive, zero, and negative values in  Figs. \ref{fig2}, \ref{fig3} and  \ref{fig4}, respectively.  
From these figures, we observe that the behavior of mass concerning horizon radius contrasts with Maxwell's case for the phantom case. Here, the correction parameter effects are significant for the small black hole. However, the effects are negligible for a large horizon radius value.

\section{First order corrections to thermodynamic potentials}\label{sec5}
\begin{figure}[htbp]
\centering
\begin{tabular}{cc} 
    \includegraphics[width=0.45\textwidth]{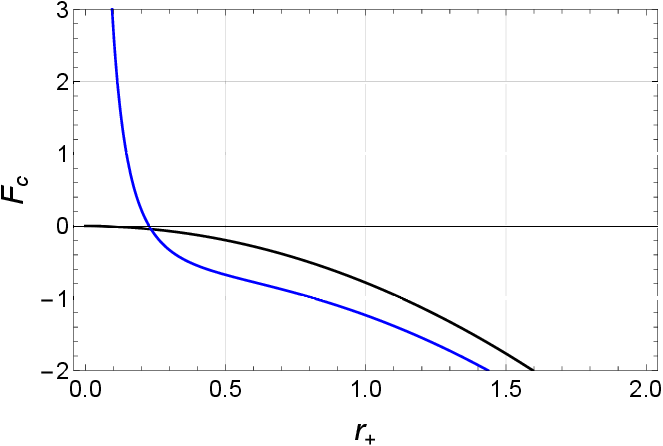}
     \includegraphics[width=0.45\linewidth]{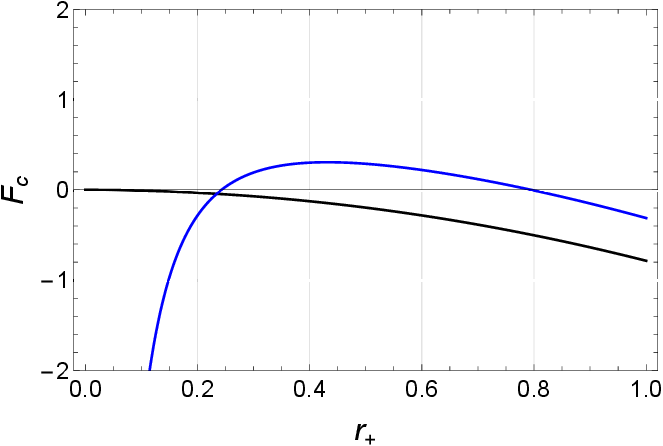}
 \end{tabular}\caption{Helmholtz free energy vs $r_{+}$ for  BTZ black hole for $\Lambda = 0.1$. The black curve represents $\alpha = 0$, and the blue curve represents $\alpha = 0.5$. Left panel: $\eta=1$ (phantom case). Right panel: $\eta=-1$ (Maxwell case).  }
    \label{fig5}
\end{figure}

The corrected Helmholtz free energy is given by 
\begin{equation}
 F_{c} = - \int S_{c} \, dT_{H},
 \end{equation}
Therefore,
\begin{eqnarray}
F_c &=& \frac{1}{4 \pi r_+} \left(
     -\pi^2 r_+^3 - 8Q^2 \alpha \eta (1 + \ln[8\pi]) 
    + 2r_+^2 \alpha \Lambda (3 + \ln[8\pi])\right. \nonumber\\
     & +& \left.\left(8 Q^2 \alpha \eta - 2r_+^2 \alpha \Lambda \right)
    \ln\left[\frac{(-4Q^2 \eta + r_+^2 \Lambda)^2}{r_+}\right]
\right).
\label{Helmholtz_btz}
\end{eqnarray}
The effects of the leading-order correction to Helmholtz free energy can be seen in Fig. \ref{fig5}.  
 From the plot, we observe that without thermal fluctuation, the behavior of Helmholtz free energy concerning horizon radius is the same and becomes more negative with horizon radius for both phantom and Maxwell cases. However, thermal fluctuation plays an essential role in small black holes as 
 Helmholtz free energy takes asymptotically large and opposite values for both cases as the size of the black hole reduces.

In the context of black holes, the Gibbs free energy is defined as 
\begin{equation}
  G_c = M_c - T_{H}S_c - QU.
  \label{Gibbs_M}
\end{equation}
\begin{eqnarray}
G_c &= &\frac{\eta q^2}{4} \ln\left( \frac{r_{+}}{l} \right) 
- \frac{\Lambda r_{+}^2}{8} 
+ \alpha \left( -\frac{\eta q^2}{2 \pi r_{+}} + \frac{3 \Lambda r_{+}}{2 \pi} \right) \nonumber \\
& -&\left( \frac{\eta q^2}{2\pi r_{+}} - \frac{\Lambda r_{+}}{2\pi} \right) 
\left( \frac{\pi r_{+}}{2} 
- \alpha \ln\left[ \frac{\pi r_{+}}{2} \left( \frac{\eta q^2}{2\pi r_{+}} - \frac{\Lambda r_{+}}{2\pi} \right)^2 \right] \right) \nonumber \\
&  -& \frac{q^2}{2} \ln\left( \frac{r_{+}}{l} \right),
\end{eqnarray}
and in terms of uncorrected entropy $S_0$ this can be expressed as 
\begin{eqnarray}
G_c &=& \eta Q^2 \ln\left( \frac{2S_0}{\pi l} \right) 
- \frac{\Lambda S_0^2}{2\pi} \nonumber \\
&& +\ \alpha \left( -\frac{\eta Q^2}{S_0} + \frac{3 \Lambda S_0}{\pi^2} \right) \nonumber \\
&& -\ \left( \frac{\eta Q^2}{S_0} - \frac{\Lambda S_0}{\pi^2} \right)
\left( S_0 - \alpha \ln\left[ S_0 \left( \frac{\eta Q^2}{S_0} - \frac{\Lambda S_0}{\pi^2} \right)^2 \right] \right) \nonumber \\
&& -\ 2 Q^2 \ln\left( \frac{2S_0}{\pi l} \right).
\label{G_corrected}
\end{eqnarray}

To do a comparative analysis, we plot expression (\ref{G_corrected}) 
for different charge and negative cosmological constant values in the phantom and Maxwell cases.
  \begin{figure}[htbp]
\centering
\begin{tabular}{cc} 
     \includegraphics[width=0.45\textwidth]{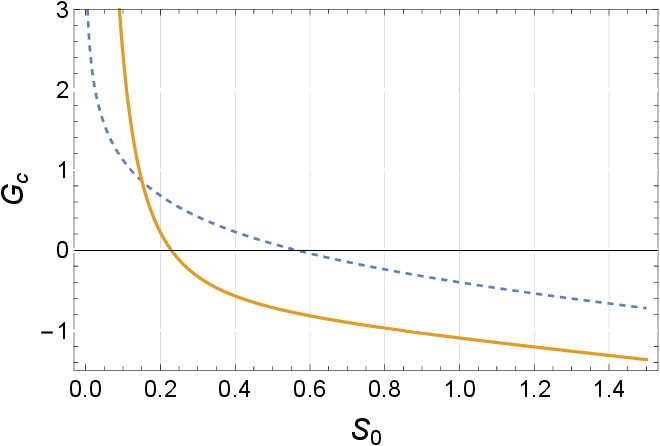} 
     \includegraphics[width=0.45\textwidth]{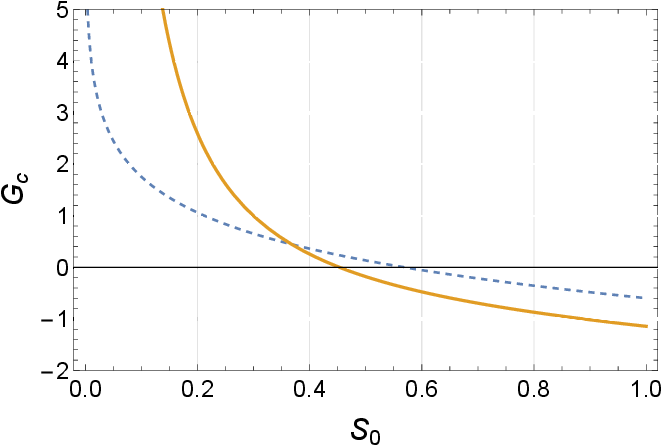}
 \end{tabular}
 \caption{$G_c$ vs $S_0$ for $\Lambda = -1$ and $\eta = 1$ (Phantom case).
 Here, $\alpha = 0$ is represented by a dashed curve, and a thick curve represents $\alpha = 0.5$. 
  Left panel:   $Q = 0.8$. Right panel:   $Q = 1.0$.}
 \label{fig6}
\end{figure} 
We observe that thermal fluctuations cause a significant decrease in the Gibbs free energy. For the phantom case, the Gibbs free energy becomes less negative as charge $Q$ increases from 0.8 to 1.0  (Fig. \ref{fig6}). Whereas, for the Maxwell case, we see that corrected Gibbs free energy increases and becomes positive for both values of charges (Fig. \ref{fig7}).     
\begin{figure}[htbp]
\centering
\begin{tabular}{cc} 
     \includegraphics[width=0.45\textwidth]{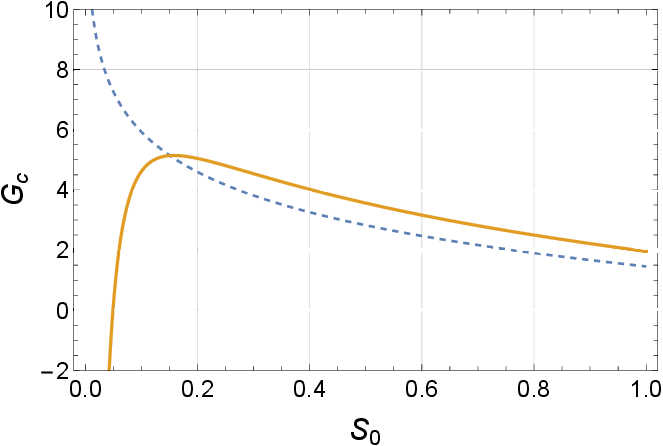} 
     \includegraphics[width=0.45\textwidth]{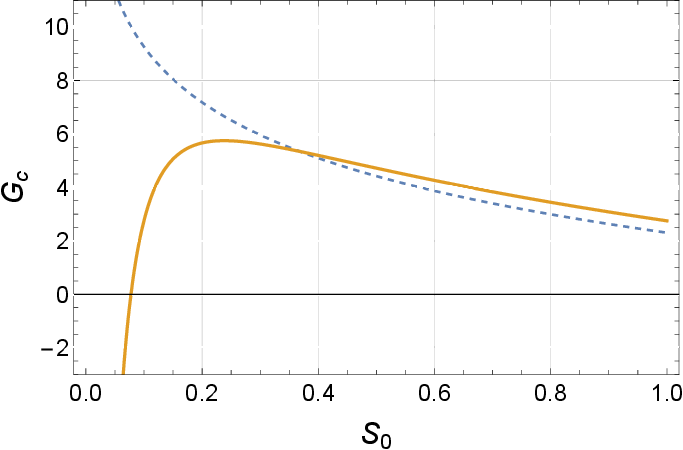}
   \end{tabular}
 \caption{$G_c$ vs $S_0$ for $\Lambda = -1$ and $\eta = -1$ (Maxwell case).
 Here, $\alpha = 0$ is represented by a dashed curve, and a thick curve represents $\alpha = 0.5$. 
  Left panel:   $Q = 0.8$. Right panel:   $Q = 1.0$.}
\label{fig7}
\end{figure} 

\begin{figure}[htbp]
\centering
\begin{tabular}{cc} 
     \includegraphics[width=0.45\textwidth]{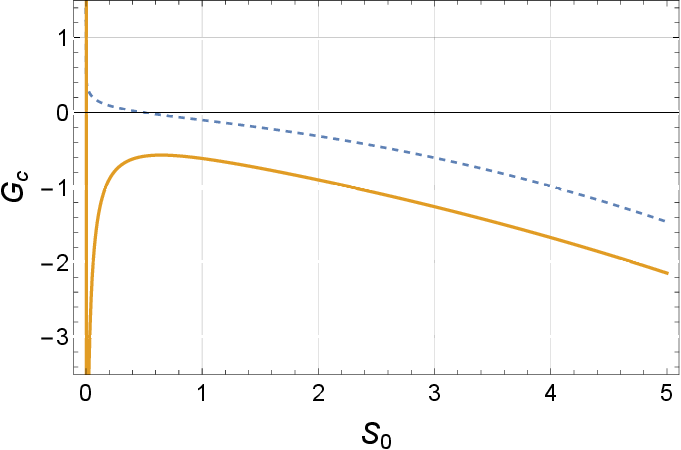} 
     \includegraphics[width=0.45\textwidth]{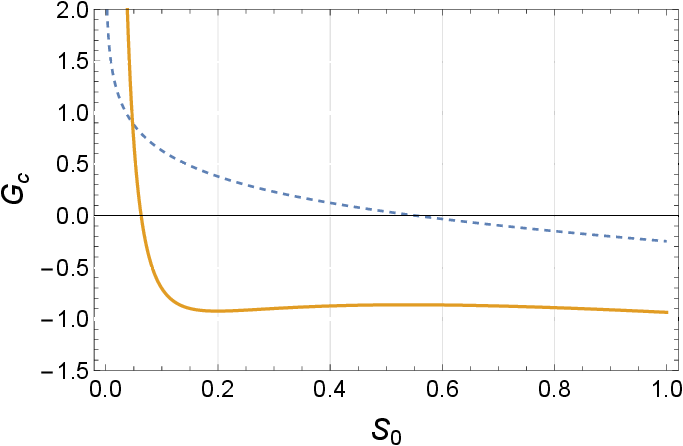}
   \end{tabular}
   \caption{$G_c$ vs $S_0$ for $\Lambda = -1$ and $\eta = 1$ (Phantom case).
 Here, $\alpha = 0$ is represented by a dashed curve, and a thick curve represents $\alpha = 0.5$. 
  Left panel:   $Q = 0.3$. Right panel:   $Q =0.6$. }
\label{fig8}
\end{figure}

\begin{figure}[htbp]
\centering
\begin{tabular}{cc} 
     \includegraphics[width=0.45\textwidth]{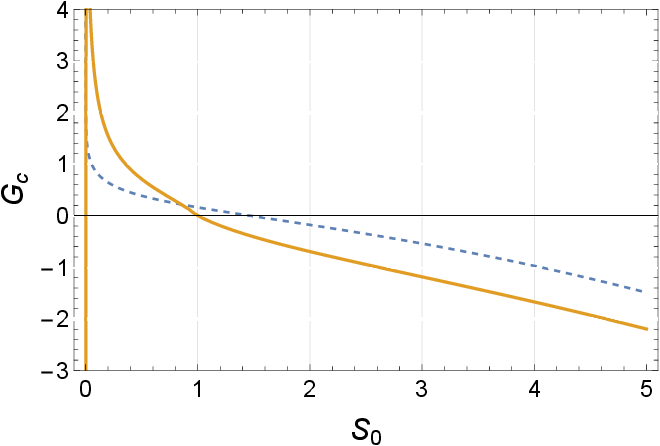} 
     \includegraphics[width=0.45\textwidth]{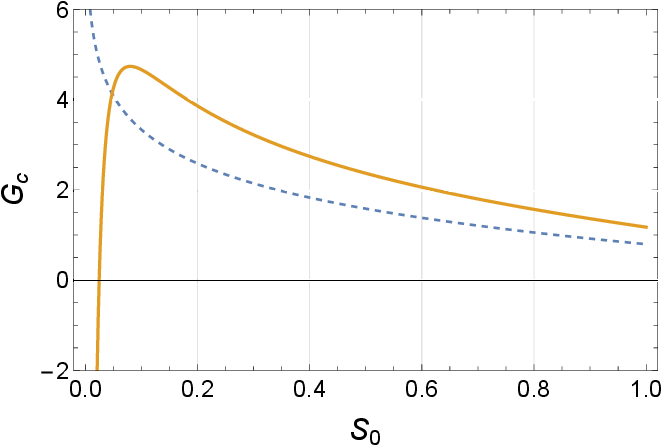}
   \end{tabular}
   \caption{$G_c$ vs $S_{0}$ for $\Lambda = -1$ and $\eta = -1$ (Maxwell case).
 Here, $\alpha = 0$ is represented by a dashed curve, and a thick curve represents $\alpha = 0.5$. 
  Left panel:   $Q = 0.3$. Right panel:   $Q = 0.6$. }
\label{fig9}
\end{figure}

 For the phantom case with $Q = 0.3$, we see that initially, the corrected Gibbs free energy decreases sharply, reaches a minimum value, exhibits a local maximum, and decreases again. On the other hand, for $Q = 0.6$ we see that there is a significant decrease of Gibbs free energy up to some finite negative value; afterwards, the Gibbs free energy behaves nearly as a constant valued function of unperturbed entropy $ S_0$  (Fig. \ref{fig8}).
 Further, for the Maxwell case with $Q = 0.3$, we observe that the corrected Gibbs free energy increases sharply concerning $ S_0$  and takes a maximum value, afterwards decreases significantly, and becomes negative. For $Q=0.6$, we inferred that the corrected Gibbs free energy increases and attains a local maximum. After that, its value decreases but remains positive (Fig. \ref{fig9}).
 
 Incidentally, we observe that the behavior of Gibbs free energy due to thermal fluctuations is opposite to that of the equilibrium value in both Maxwell and phantom field background cases.

 {We also see from the figure  \ref{fig6} that Gibbs free energy is positive for small values of unperturbed entropy, i.e, for small black holes, and Gibbs free energy is negative for larger black holes. A negative Gibbs free energy indicates that a black hole is thermodynamically favored and globally stable in a given ensemble. Conversely, a positive value suggests the black hole is unstable and may undergo phase transitions to acquire stability. For the Maxwell case from figure \ref{fig7}, we observe that the corrected Gibbs free energy increases from a negative value to a positive value for both charges, suggesting that the black hole may undergo a phase transition to gain stability. From Fig.  \ref{fig8}  we see that the Gibbs free energy is negative for large black holes. Large black holes are globally stable for charges $Q = 0.3$ and $Q = 0.6$, and are less susceptible to undergoing phase transitions. Also, for $Q = 0.3$ in Fig.  \ref{fig9}, we see that the Gibbs free energy is negative for larger black holes, suggesting that large black holes are thermodynamically favored in this case. However, globally, we find that for $Q = 0.6$, the Gibbs free energy is positive for large black holes. This implies that, for this case, large black holes are not globally stable and might undergo a phase transition to acquire thermodynamic stability.    }

\section{Corrected Specific heat and Stability}\label{sec6}

The heat capacity or specific heat for the phantom BTZ black hole is given by
\begin{equation}
 C_{Q} = T_{H} \left(\frac{\partial S_{c}}{\partial T_{H}} \right)_{Q} . 
\end{equation}
Now, by plugging the values of $T_{H}$ and $S_{c}$ from  Eqs. (\ref{BTZ_Temp}) and   (\ref{corr}), respectively, in the above relation,  we get
\begin{equation}
C_{Q} = \left(\frac{\eta q^2}{2 \pi r_{+}} - \frac{\Lambda r_{+}}{2 \pi}\right) \left(\frac{\pi}{2} + \frac{\alpha (q^2 \eta + 3r^2_{+}\Lambda)}{q^2 r_{+} \eta - r^3_{+} \Lambda}\right) \left(\frac{-2 \pi r^2_{+}}{q^2 \eta + r^2_{+} \Lambda}\right) ,
\label{C_r}
\end{equation}
and in terms of uncorrected entropy $S_{0}$ this can be written as:
\begin{equation}
C_{Q} =  \left(\frac{\eta Q^2}{S_0} - \frac{\Lambda S_0}{\pi^2}\right) \left( \frac{\pi}{2} + \frac{\alpha \pi (4Q^2 \eta \pi^2 + 12S_0^2 \Lambda)}{8Q^2 S_0 \eta \pi^2 - 8S_0^3 \Lambda}\right) \left(\frac{-8S_0^2 \pi}{4Q^2 \eta \pi^2 + 4S_0^2 \Lambda}\right).   
\end{equation}
 We study the nature of its specific heat to analyze the stability of black holes. We can estimate whether a black hole undergoes a phase transition from the nature of specific heat.
The positive value of specific heat confirms that the system is against the phase transition. In contrast, the negative value of specific heat indicates the system's instability.

\begin{figure}[htbp]
\centering
\begin{tabular}{cc} 
     \includegraphics[width=0.45\textwidth]{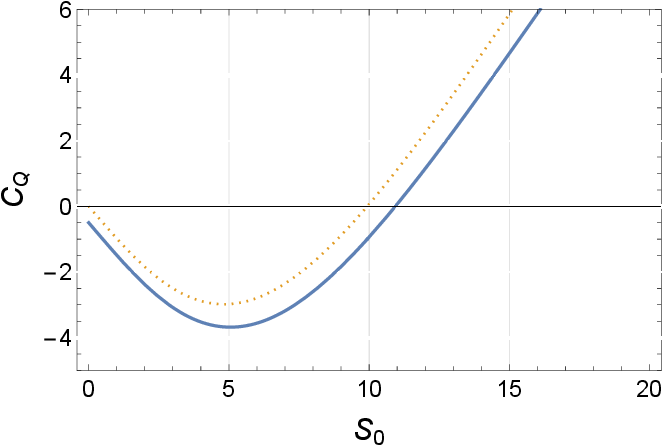}  
     \includegraphics[width=0.45\textwidth]{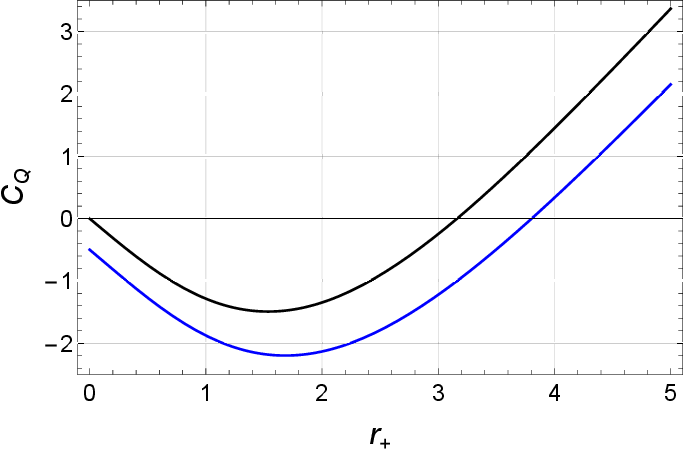}
     \end{tabular}
     \caption{Corrected specific heat $(C_Q)$ vs entropy ($S_0$).  Left panel:    $\Lambda = 0.1$ and $\eta = 1$ (phantom case). Here, $\alpha = 0$ is represented by a dotted curve, and a thick curve represents $\alpha = 0.5$. Right panel: $C_Q$ vs event horizon radius $r_+$ for  $\Lambda=0.1$ and $\eta = 1$ (Phantom case). The black curve represents $\alpha = 0$, and the blue curve represents $\alpha = 0.5$.}
 \label{fig12}
\end{figure}

 We see from the equation (\ref{C_r}) that the expression for specific heat is the same for both phantom ($\Lambda =0.1$ and $\eta = 1$) and Maxwell case ($\Lambda = -0.1$ and $\eta = -1$).
 In Fig. \ref{fig12}, we see the behavior of specific heat of a phantom BTZ black hole ( for $\Lambda = 0.1$ and $\eta = 1$), which is plotted concerning unperturbed entropy $S_0$ and event horizon radius $r_+$.
 
 These two diagrams suggest that the phase transition occurs for such a black hole. The specific heat is negative for small black holes, which indicates that small black holes are thermodynamically unstable. On the other hand, the specific heat is always positive for larger black holes, which means these black holes are in a stable phase.
We see that the specific heat varies from negative to increasingly positive values.

\section{$P-V$ criticality}\label{sec7}
The pressure of the phantom BTZ black hole can be given as:
\begin{equation}
  P_c = -\left(\frac{\partial F_c}{\partial V}\right)_{T_{H}},
  \label{P_btz}
\end{equation}
where $F_{c}$ denotes the corrected Helmholtz free energy given by equation (\ref{Helmholtz_btz}), and volume $V = \pi r^2_{+}$.

Therefore, the pressure of the phantom BTZ black hole by using Eq.  (\ref{Helmholtz_btz}) and (\ref{P_btz}) can be given as:
\begin{equation}
 P_c = \frac{1}{4 \sqrt{\pi V^3} } \left[\pi^{\frac{1}{2}} V^{\frac{3}{2}}
+ \alpha \left(4Q^2\eta + \frac{V}{\pi} \Lambda\right) \ln\left[\frac{1}{8}\sqrt{\frac{1}{\pi V}}\left(-4Q^2\eta + \frac{V}{\pi} \Lambda\right)^2 \right]\right].
 \label{P_c}
\end{equation}
Now, the critical points can be obtained by the derivative of pressure concerning volume $V$ of the phantom BTZ black hole, and are given by
\begin{eqnarray}
\left(\frac{\partial P_c}{\partial V}\right)_{T_H = T_c} &= &\frac{\alpha}{8 \pi^{3/2} V^{5/2} (4 \pi Q^2 \eta - V \Lambda)}
\left[
-8 \pi Q^2 V \eta \Lambda (2 + \ln [8 \pi]) \right.\nonumber\\
&-& \left.V^2 \Lambda^2 (3 + \ln [8 \pi]) 
+ 16 \pi^2 Q^4 \eta^2 (-1 + 3 \ln [8 \pi])\right.\nonumber \\
&-&\left. (4 \pi Q^2 \eta - V \Lambda) (12 \pi Q^2 \eta + V \Lambda)  \ln \left[ \frac{(-4 \pi Q^2 \eta + V \Lambda)^2}{\pi^{3/2} \sqrt{V}}\right]
\right] = 0.
\label{P_derivative}
\end{eqnarray}
The critical volume of a phantom BTZ black hole can be calculated by numerically solving Eq.  (\ref{P_derivative}). Here, we take the parameters for numerical computation as follows: $\eta = 1$, $\Lambda = 0.1$, $\alpha = 0.5$, and $Q =1$. Therefore, we get the value of critical volume $V_{c}$ from the equation (\ref{P_derivative}) as $V_{c} = 2.43966$.  
Critical pressure can be calculated from Eq.  (\ref{P_c}) and is given by $P_c = 0.222503$.

Therefore, the critical temperature can be calculated from 
 (\ref{BTZ_Temp}) and is given by $T_c = 0.708396$.

Critical compressibility factor ($Z_{c}$) is given by
\begin{equation}
 Z_{c} = \frac{P_{c} V_{c}}{T_{c}} = \frac{(0.222503) (2.43966)}{0.708396} = 0.76628279 .   
\end{equation}
The standard value of $Z_{c}$ for Van der Waals fluid is $\frac{3}{8}$ (0.375), and the value of $Z_{c}$ computed for our black hole system, treating it as a Van der Waals fluid, is 0.766. Therefore, our phantom BTZ black hole system differs slightly from standard Van der Waals fluid behavior.

\section{Concluding remarks}\label{sec8}
 
In this work, we have investigated the thermodynamics of phantom BTZ black holes by incorporating leading-order quantum corrections arising from small statistical fluctuations around the equilibrium of our black hole system. Starting from a modified action in three-dimensional spacetime that includes coupling with a Maxwell or phantom field, we considered the corresponding field equations and BTZ black hole solution for the metric incorporating the phantom field to study the effect of thermal fluctuations on the thermodynamics of small phantom BTZ black holes. 

We have computed the corrected entropy using the steepest descent method and expressed it in terms of the leading entropy and Hawking temperature. Furthermore, using standard thermodynamic relations, we have derived the corrected mass, Helmholtz free energy, specific heat, and Gibbs free energy. These corrections have revealed significant deviations from classical results, especially in the small black hole regime where quantum effects become prominent.

Through graphical analysis, we have shown that the corrected entropy can become negative for sufficiently small black holes, highlighting potential limitations in thermodynamic stability under quantum corrections.{The negative value of the corrected entropy is unphysical here; however, we also saw that corrected entropy gains more positive value and is an increasing function of horizon radius for large black holes. We also observed that small statistical thermal fluctuations do not significantly affect the entropy behavior for huge black holes.} Similarly, we have demonstrated that thermal fluctuations have a pronounced effect on all thermodynamic potentials for small black holes. At the same time, their influence diminishes for larger black holes.

We have also analysed the stability of phantom BTZ black holes by calculating their specific heat. We have found that a phase transition occurs for such black holes.
For small phantom BTZ black holes, the specific heat takes negative values, which implies that small black holes are in a thermodynamically unstable phase. At the same time, the specific heat is found to be increasingly positive for larger black holes, from which we inferred that these black holes are in a stable phase.
Treating our phantom BTZ black hole system as a Van der Waals fluid, we have also computed its critical points $P_{c}$, $V_{c}$, $T_{c}$, and critical compressibility factor $Z_{c}$.

From the computed value of critical compressibility factor $Z_{c}$, we inferred that our black hole system slightly differs from the standard Van der Waals fluid behavior.

{The fundamental physical consequence of this work is that through the computation of corrected specific heat, phase transition phenomena and stability of phantom BTZ black holes have been analyzed for the first time. Also, by analyzing P-V criticality in this work, we found that phantom BTZ black holes deviate slightly from the standard Van der Waals fluid. Possibly, this is due to the incorporation of phantom fields around the BTZ black holes.     }

Overall, this study has provided a comprehensive understanding of how small statistical thermal fluctuations modify the thermodynamics of phantom BTZ black holes and has emphasized the importance of including such corrections in any realistic model of black hole thermodynamics in $(2+1)$ dimensions.

\end{document}